\documentclass[%
 amsmath,amssymb,
 aps,
]{revtex4-2}

\usepackage{graphicx}
\usepackage{dcolumn}
\usepackage{bm}

\begin{document}

\preprint{ParkerLab/3000}

\title{Spin Susceptibility Above the Superfluid Onset in Ultracold Fermi Gases}

\author{Yun Long}
\author{Feng Xiong}
\author{Colin V. Parker}
\affiliation{School of Physics, Georgia Institute of Technology, Atlanta, GA 30332}

\date{\today}

\begin{abstract}

Ultracold atomic Fermi gases can be tuned to interact strongly, where they display spectroscopic signatures above the superfluid transition reminiscent of the pseudogap in cuprates. However, the extent of the analogy can be questioned, since thermodynamic quantities in the low temperature spin-imbalanced normal state can be described successfully using Fermi liquid theory. Here we present spin susceptibility measurements across the interaction strength-temperature phase diagram using a novel radiofrequency technique with ultracold $^6\textrm{Li}$ gases. For all significant interaction strengths and at all temperatures we find the spin susceptibility is reduced compared with the equivalent value for a non-interacting Fermi gas, with the low temperature results consistent with previous studies.  However, our measurements extend to higher temperatures, where we find that the reduction persists consistently with a mean-field scenario. At unitarity, we can use the local density approximation to extract the spin susceptibility for the uniform gas, which is well described by mean-field models at temperatures from the superfluid transition to the Fermi temperature.
\end{abstract}

\maketitle

\newcommand{\ket}[1]{\left|#1\right\rangle}

In the study of strongly interacting quantum systems, the BEC-BCS crossover is a simple and experimentally realizable model\cite{nozieres_bose_1985,giorgini_theory_2008,randeria_crossover_2014} with implications for a variety of physical systems, such as the high-$T_c$ cuprates\cite{chen_bcsbec_2005} and neutron matter\cite{margueron_bcs-bec_2007,strinati_bcsbec_2018}. In this model, a two-component Fermi gas with contact interactions has an attractive contact interaction of varying strength. When the interaction is weak, the Bardeen-Cooper-Schrieffer (BCS) state forms, while for strong interactions the fermions form composite (bosonic) molecules, which then condense into a Bose-Einstein condensate (BEC). The crossover occurs as these two states connect to one and other by tuning the interaction strength, parameterized by $(k_F a)^{-1}$, where $k_F$ is the Fermi wavevector and $a$ is the s-wave scattering length. A fundamental question that emerges from this model is the degree to which it embodies something universal about strongly interacting fermions, that is, which strongly interacting fermionic systems can be approximately mapped to an effective theory falling somewhere along the crossover. In particular, for the cuprates, the phenomenon of depressed density of states above the transition temperature known as the pseudogap\cite{timusk_pseudogap_1999,mueller_review_2017} has been suggested to be due to a ``preformed pairs'' state analogous to the BEC-BCS crossover\cite{chen_bcsbec_2005}. Although one does not expect the complete, complex phenomenology of cuprates in the BEC-BCS crossover, it remains an open question whether the cuprate pseudogap derives fundamentally from a strongly interacting pairing mechanism. If so, one would expect an analog pseudogap in ultracold gases.

In light of these possible similarities, an important project is to compare the measured properties between ultracold gases at the BEC-BCS crossover with their equivalent in materials. For spectroscopic properties, one can compare angle-resolved photoemission spectroscopy (ARPES) measurements of the cuprates\cite{ding_spectroscopic_1996,timusk_pseudogap_1999,damascelli_angle-resolved_2003,vishik_photoemission_2018} with their ultracold gas analogs\cite{stewart_using_2008,gaebler_observation_2010,sagi_breakdown_2015}. However, one expects spectroscopic measurements to be strongly influenced by the nature of the particles' dispersion curve, which is parabolic for ultracold gases but significantly not so in materials\cite{norman_phenomenological_1995}, which in addition can have surface effects. Bulk thermodynamic measurements\cite{luo_measurement_2007,ho_universal_2004,Valtolina2017} are an alternative, as they can be deduced from the equation of state (EoS) of a trapped atomic gas\cite{zwierlein_direct_2006,chevy_universal_2006,shin_phase_2008,shin_determination_2008,nascimbene_exploring_2010,navon_equation_2010}. An appealing property for comparisons is the spin susceptibility. This property is easily calculated for the non-interacting Fermi gas, which provides a natural scale, and is directly measurable in materials, for example using NMR Knight shift\cite{walstedt_63cu_1990,takigawa_cu_1991,timusk_pseudogap_1999}. In these NMR measurements, a clear decrease of the spin susceptibility is observed as the sample is cooled below the pseudogap onset temperature, $T^*$, associated with a reduction in the density of states at the Fermi energy. In ultracold atomic gases, the spin susceptibility has been determined from the EoS\cite{nascimbene_fermi-liquid_2011}, from out-of-equilibrium dynamics\cite{sommer_universal_2011,Valtolina2017} and measured using speckle-field imaging\cite{sanner_speckle_2011}. These studies found that the spin susceptibility near the onset of superfluidity is significantly reduced from the expected value for a non-interacting gas, but did not establish the temperature dependence. Such reductions are consistent with any combination of a temperature-independent mean-field reduction due to interactions, and a pseudogap setting in below a specific $T^*$. Here we present the first comprehensive study of spin susceptibility over the entire interaction-temperature phase diagram, demonstrating that it is consistent with a mean-field origin and any pseudogap effects are small, in particular in comparison with those seen in cuprates.

\begin{figure}[t]
\includegraphics[width=0.5\textwidth]{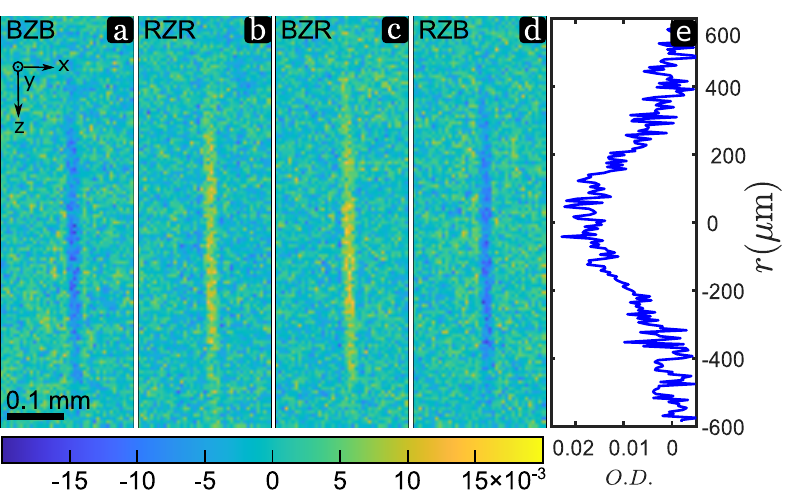}
\caption{\label{fig:meas_rf} Four closely related experiments with the same atomic cloud preparation, but different RF procedures. (a)-(d) The experiments described as BZB, RZR, BZR, RZB respectively, in the text where B and R represent initial or final blue or red detuning and Z represents a 2 s on resonant hold time. The RF procedure generates a spin difference leading to positive signal in (b) and (c) and negative signal in (a) and (d). (e). The net differential signal (-BZB+RZR+BZR-RZB)/4 as a 1D profile along the long ($z$) axis of the trap, after summing over the $x$ direction with a $29\,\mu\textrm{m}$ wide region centered on the atomic cloud (the $y$ direction is naturally summed by the imaging). The data comes from a unitary gas at $210\,\textrm{nK}$ ($T/T_F^{t} = 0.27$).}
\end{figure}

We developed a novel method to measure the spin susceptibility by radiofrequency (RF) dressing. With a resonant RF driving between two hyperfine states, in the interaction picture, a chemical potential difference $\Delta\mu = \mu_+ - \mu_- =\hbar\Omega$ between the two dressed states is created. This can be thought of as an effective Zeeman field, and we can extract the susceptibility from the dressed states' number difference after equilibrium is reached.
Labeling the hyperfine states of $^{6}\textrm{Li}$ with $\ket{1}$ from the bottom, we use a mixture of states $\ket{2}$ and $\ket{3}$, which are adiabatically connected to the $\ket{m_J,m_I} = \ket{-1/2,0}$ and $\ket{-1/2,-1}$ states, respectively, in the limit of high field. We begin by preparing this mixture using standard ultracold atom methods, yielding a Fermi gas with $20,000 - 100,000$ atoms in each spin state at temperatures ranging from several times the trap Fermi temperature $T_F^t$ down to $0.15\cdot T_F^t$\cite{long_all-optical_2018}. The initial mixture is spin imbalanced with a typical majority:minority ratio of 2:1, and the gas is held in a single-beam optical trap, with confinement along the beam axis provided by the magnetic field. Typical trapping frequencies are $30\textrm{ Hz}$ along the beam axis and $1-2\textrm{ kHz}$ in the perpendicular directions. We then expose the gas to RF radiation on the $\ket{2} \Leftrightarrow \ket{3}$ transition. The radiation is initially $100\,\textrm{kHz}$ detuned and then adiabatically ramped onto resonance in $47\,\textrm{ms}$, mapping the spin imbalance from the initial basis of $\ket{2}$ and $\ket{3}$ states into an imbalance in the RF-dressed basis, which we denote by $\ket{+}$ and $\ket{-}$. In the rotating wave approximation (RWA), these states have energies of $+h\Omega/2$ and $-h\Omega/2$, respectively, where $\Omega$ is the Rabi frequency. In our experiment, $\Omega=1.4\,\textrm{kHz}$ which is small compared to other energy scales and gives a linear response (see supplementary materials). Once the RF radiation is on resonance, we allow the sample to reach equilibrium for a holding time of typically $2\,\textrm{s}$. The gas is held in a magnetic field gradient, which provides a large scalar force together with a small spin-dependent force. The spin dependent force is small (of the order $\textrm{nK}/\mu\textrm{m}$) because the magnetic moment of the two states is very nearly equal. However, the spin-dependent potential gradient associated with this force is sufficient to allow the dressed state populations to exchange and reach thermodynamic equilibrium (see supplementary materials). Following the hold period, we adiabatically ramp the radiation to a $100\,\textrm{kHz}$ detuning, which maps the imbalance back into the $\ket{2}$-$\ket{3}$ basis, where we image the sample {\it in-situ} using phase-contrast imaging.

We choose the sign of the initial and final detuning in a specific way to avoid possible bias from slight imaging frequency offsets and/or residual initial spin imbalance. We perform the experiment will all four possible signs of initial and final detuning, using otherwise identical procedure. To simplify the description, we use  ``R'' and ``B'' to denote initial or final detuning which is negative (red) or positive (blue) and ``Z'' to denote zero detuning, or resonant RF. Thus a ``BZB'' experiment consists of ramping the RF frequency from above resonance onto the resonance, holding for 2 s, and ramping again to higher frequency. Phase contrast images of the differential spin density from a typical experimental run after applying our enhanced principle component algorithm\cite{xiong_enhanced_2020} are shown in Fig. \ref{fig:meas_rf} (a) - (d), with positive signal corresponding to an excess density of state $\ket{2}$.  The $-\textrm{BZB}+\textrm{RZR}$ and $\textrm{BZR}-\textrm{RZB}$ differences are insensitive to any hypothetical initial imbalances that persist through the experiment, since $\textrm{BZB}$/$\textrm{RZR}$ and $\textrm{BZR}$/$\textrm{RZB}$ are mirrored pairs from the perspective of Landau-Zener sweeps, which respectively leave the spin population unchanged and invert the imbalance, in the limit of no relaxation. Similarly the $-\textrm{BZB}+\textrm{BZR}$ and $\textrm{RZR}-\textrm{RZB}$ differences are insensitive to imaging offset, because the dressed spin imbalance prior to the final ramp is mapped onto opposite final states. Hence the quantity $-\textrm{BZB}+\textrm{RZR}+\textrm{BZR}-\textrm{RZB}$ corrects for both possibilities. Fig. \ref{fig:meas_rf} (e) shows the row-summed differential signal generated by our RF method. These data will form the basis for our analysis. It is worth noting that our method for generating the imbalance is distinct from our imaging method, and that therefore any method of spin-selective imaging could be employed, such as a quantum gas microscope, MOT recapture, or resonant ionization detection.

\begin{figure}
\includegraphics[width=0.5\textwidth]{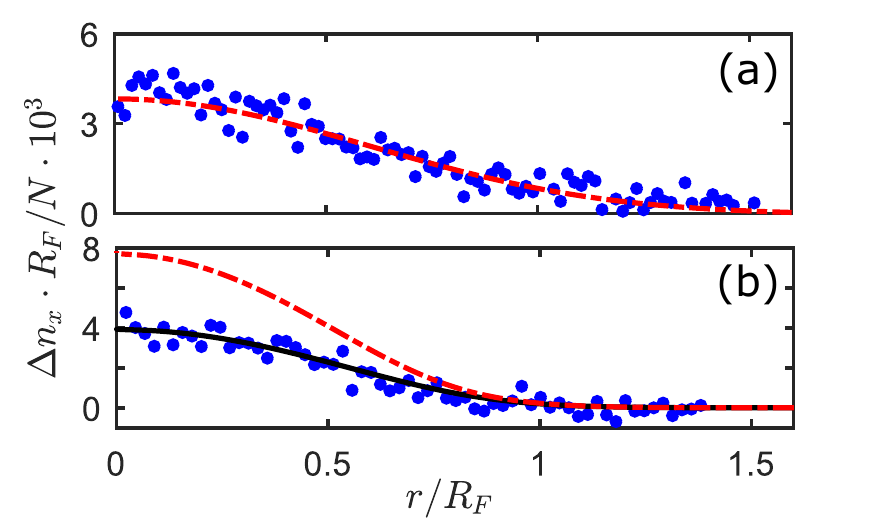}
\caption{\label{fig:spin_diff} Differential column density measured at (a) $40.0\,\textrm{mT}$ where interaction is close to zero at $513\,\textrm{nK}$ ($T/T_F^{t} = 0.63$) and (b) at $81.4\,\textrm{mT}$ near the unitary point at $210\,\textrm{nK}$ ($T/T_F^{t} = 0.27$). The red dashed lines in (a) and (b) show the non-interacting susceptibility for a gas with the same density profile. In (b), the black solid line is from our mean-field model.}
\end{figure}

By calibrating with the total atom number in each spin state, we can determine the long axis differential axial density profile $\Delta{n}_z$. We define an orthogonal coordinate system where $\hat{z}$ runs along the trapping beam axis, $\hat{y}$ is in the direction of gravity, the magnetic field, and the magnetic gradient, and $\hat{x}$ is the remaining direction. The axial density represents the density integrated over $x$ and $y$ therefore. We present this for a weakly interacting gas and a unitary gas in Fig. \ref{fig:spin_diff} in the scaled form $\Delta{n}_z\cdot\textrm{R}_F/N$,  where N is the total atomic number, $R_F=\sqrt{2E_F^{t}/m\Omega^2_z}$ is the Thomas-Fermi radius, $E_F^{t} = k_B T_F^{t}$ is the (trap) Fermi energy and $\Omega_z$ is the long-axis trap frequency. For comparison to theoretical models, we acquire the long axis total axial density profile $n_z$, and use this information together with the local density approximation (LDA) and a published equation of state (EoS) measurement at unitarity\cite{Ku563} to determine the three dimensional density profile in our trap. Away from unitarity we use a phenomenological fit based on a polylogarithm, which yields similar results (see supplementary information). From the density profile, we can compute the expected spin difference from a susceptibility model and integrate to generate the expected spin difference. We do this for two models, the ideal Fermi gas model, which describes the weakly interacting data well but not the unitary data, and a mean field model described below.

\begin{figure}
\includegraphics[width=0.5\textwidth]{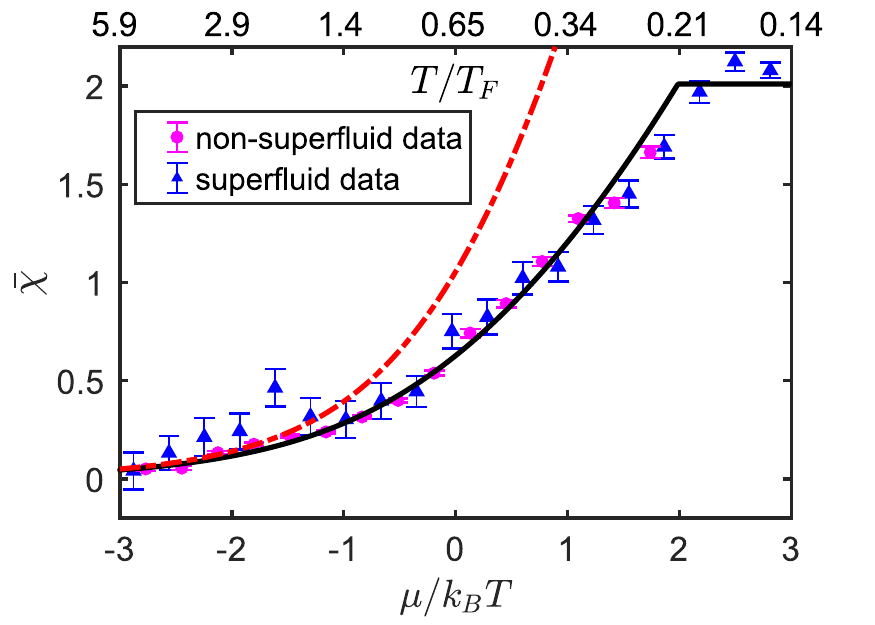}
\caption{\label{fig:mean_field} Dimensionless integrated susceptibility $\bar{\chi}$. The black solid line is the mean field prediction with $\alpha=1.61$, the red dashed line is the non-interacting susceptibility for a Fermi gas with the same density profile. Purple solid circles are a binned average of non-superfluid samples, while the blue solid triangles are taken at a temperature at which the superfluid phase appears at the trap center, and have been scaled to correct for a parasitic evaporative effect. Error bars are derived from the standard deviation of the measurements.}
\end{figure}

We model the spin susceptibility of the system, which is given by
\begin{align}
    \chi^{-1} = \partial H/\partial M = \left(\partial^2 F/\partial M^2\right)_{T,V,N},
\end{align}
where $F$ is the Helmholtz free energy of the system, $H = \mu_\uparrow-\mu_\downarrow$ is the analog of magnetic field, and $M = N_\uparrow-N_\downarrow$ is the magnetization. In a mean-field picture, the interaction simply adds a temperature-independent term $F_I = \alpha M^2/(2\chi_\textrm{NI}^0)$ to the free energy, where $\chi_\textrm{NI}^0$ is the non-interacting susceptibility at zero temperature (in Fermi-liquid theory $\alpha$ is proportional to the parameter $F_0^a$). A more complete Fermi-liquid theory would also add an effective mass, but the effective mass correction is small for strongly interacting Fermi gases\cite{nascimbene_fermi-liquid_2011}. This then results in a susceptibility $\chi^{-1}(T) = \chi_\textrm{NI}^{-1}(T) + \alpha{\chi^0_{NI}}^{-1}$, where $\chi_\textrm{NI}(T)$ is the non-interacting spin susceptibility at temperature T.

At unitarity, where the entire trap has the same interaction parameter, we can scale the data into a dimensionless form that is independent of trapping conditions, and use the LDA to express our measurement purely in terms of the local properties of a uniform gas. In particular we consider the dimensionless integrated susceptibility given by:
\begin{align}
    \bar{\chi} = \int_{-\infty}^{\mu} \chi(\mu') \lambda_{dB}^3 d\mu',
\end{align}
where $\lambda_{dB} = \sqrt{2\pi\hbar^2/mk_BT}$ is the thermal de Broglie wavelength.
This is shown in Fig. \ref{fig:mean_field}. For lower temperature samples, at the center of the trap the data shows a plateau (blue points in Fig. \ref{fig:mean_field}). This is expected for a superfluid for which the spin susceptibility has vanished. Indeed, the onset occurs where the local $T/T_F = 0.21$, below a theoretical estimate\cite{perali_bcs-bec_2004}, but slightly above other experiments \cite{burovski_critical_2008,magierski_onset_2011}. When the sample is partially superfluid, there are significant parasitic effects that tend to balance the spins as the sample evaporates, meaning that the absolute calibration of susceptibility is no longer possible and the calibration has to be done based on overlap with totally non-superfluid samples (see Supplementary Information). Taking all the unitary data together, over a large range of temperature down to the superfluid transition, we see significant reduction compared with the non-interacting expectation. However, we are able to describe this reduction over the whole temperature range using the mean-field model. Consistent with previous results\cite{sanner_speckle_2011} and theory\cite{tajima_uniform_2014,tajima_spin_2017}, we find the susceptibility at the onset of superfluidity to be about $33\pm3\,\%$ of the non-interacting trap-averaged value when $(k_F a)^{-1} = 0$. Extrapolating our model to the lowest temperatures the ratio with a uniform non-interacting gas would be $38\pm 1\%$, slightly less than the value of approximately $50\%$ for a model calibrated to data extrapolated from spin-imbalanced samples at low temperature\cite{nascimbene_fermi-liquid_2011}. The uncertainty in these figures is purely statistical, while any of the foreseeable systematic effects would come from parasitic equilibration or insufficient equilibration time and cause the measured susceptibility to be too low. We believe the systematic effects are no larger than 10\% of the measured signal, so 4\% of the non-interacting value (see supplementary information). Alternatively, we can apply a phenomenological psuedogap model to our data, described in the Supplementary Information, which yields a modest pseudogap that reduces the susceptibility compared to the mean field value by $28\,\%$. In either event, our key result is that the (inverse) susceptibility is different from the non-interacting value by a comparable amount at low temperature and high temperature (well above $T_c$ and into the region where $\mu < 0$). Since in the latter case there is no expectation of pairing because there is no Fermi surface and free atoms cannot be bound, the high temperature deviation from the non-interacting value is very likely to be from a mean-field mechanism. That this effect is comparable at low temperatures strongly supports the hypothesis of no, or at least very little, pairing above $T_c$, with the pairing contribution to reduced susceptibility being smaller than the mean-field one.

\begin{figure}
\includegraphics[width=0.5\textwidth]{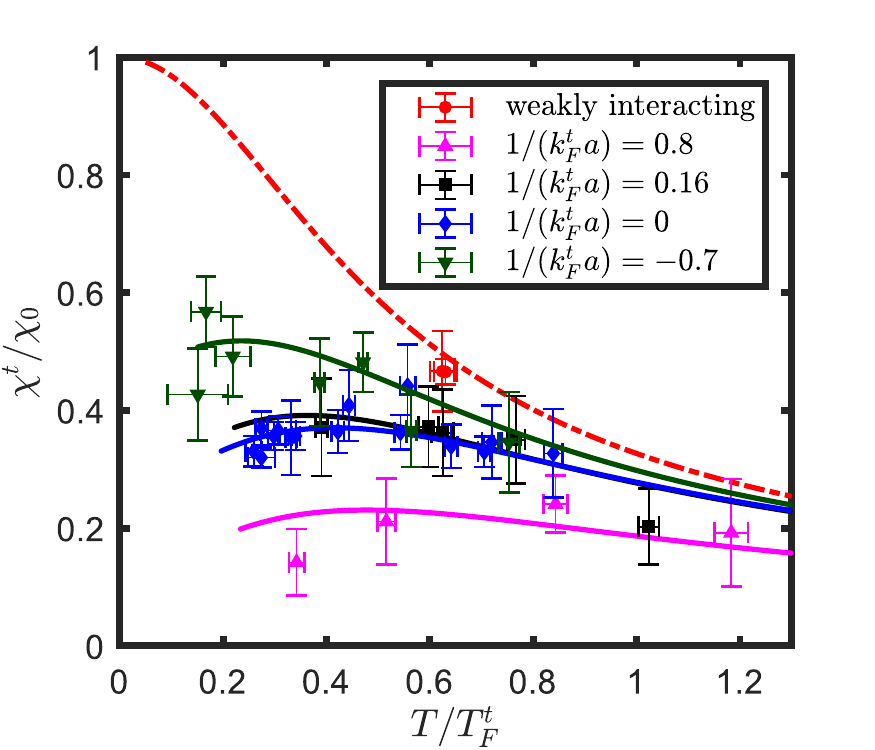}
\caption{\label{fig:susceptibility} The trap-averaged spin susceptibility normalized to the zero-temperature non-interacting value for weak interactions and across the BEC-BCS crossover. The red dotted line shows the susceptibility of a non-interacting Fermi gas as a function of temperature. Error bars are derived from the standard deviation of many shots (for susceptibility) and from fitting uncertainty (for temperature). Solid curves are calculated using the measured density distribution and our mean-field model under the local density approximation, with values of the fit parameter $\alpha$ given by $3.5$, $1.5$, $1.6$, and $0.9$ from top to bottom.}
\end{figure}

On the contrary, decreasing spin susceptibility with cooling is observed above $T_c$ in the cuprates\cite{walstedt_63cu_1990,takigawa_cu_1991}, and is also expected in ``preformed pairs'' scenarios\cite{chen_bcsbec_2005}. In order to evaluate the possibility of this scenario, even away from unitarity, we have performed these measurements throughout the phase diagram of the BEC-BCS crossover. Fig. \ref{fig:susceptibility} summarizes our main result from this perspective. We show trap-averaged measurements of the spin susceptibility, normalized to the value for a trapped non-interacting gas with the same atom number at zero temperature (to facilitate comparison with Ref. \onlinecite{tajima_spin_2017}). We characterize the temperature by the ratio $T/T^t_F$, where $k_BT^t_F=\hbar^3\omega_x\omega_y\omega_z(3N)^{1/3}$. Similarly, we characterize the interaction strength as $(k^t_Fa)^{-1}$, where $k^t_F={(2mk_BT^t_F)}^{1/2}$. The solid curves are calculated with our mean-field model and take into account the experimental density profiles, which leads to slight non-monotonic behavior for trap-averaged susceptibility\cite{tajima_spin_2017}. This model agrees well with our data throughout the phase diagram (a different mean field parameter is used for each value of the interaction strength). Our data are also consistent with calculations\cite{tajima_spin_2017} based on a pairing fluctuation scenario throughout the whole temperature region. In this sense, small pairing fluctuations cannot be ruled out, but the mean-field model is still a largely adequate description near unitarity in the normal state\cite{nascimbene_fermi-liquid_2011}.  This is somewhat surprising considering that spectroscopic experiments\cite{sagi_breakdown_2015} using potassium find a breakdown of Fermi liquid theory when $(ka)^{-1}\geq0.28\pm0.02$. Further to the BEC side, where $(k^t_Fa)^{-1}=0.8$, the susceptibility is drastically reduced at all temperatures. However, even there, a mean field model captures the temperature dependence of spin susceptibility until very near the superfluid transition.

In conclusion, we surveyed the spin susceptibility of strongly interacting $^{6}\textrm{Li}$ gases from the BCS side to the BEC side and from high temperatures down to the superfluid transition temperature. Somewhat surprisingly, the temperature dependence of the spin susceptibility can be modeled reasonably throughout the phase diagram using a mean-field model. Small deviations from the effects of pairing fluctuations are possible, although the present statistical uncertainty prevents a rejection of the mean field model. We note that future experiments could likely reduce the uncertainty considerably by increasing the fraction of imaging photons captured, and planned experiments imaging uniform density regions rather than integrating along the imaging axis could resolve the region just above Tc more easily. Notwithstanding, we can rule out any type of pseudogap with comparable magnitude to that seen in the high-$T_c$ cuprates, where reductions on the order of a factor of three are seen in the Knight shift\cite{timusk_pseudogap_1999,takigawa_cu_1991} with reducing temperature. The closest analogy this has in the BEC-BCS crossover of ultracold Fermi gases is the far BEC side, where there may be a significant percentage-wise reduction in the already-small susceptibility at 1.5-2 times $T_c$. All of this sits in contrast to spectroscopic evidence showing that even on the near BEC side coherent excitations are absent, meaning that coherence is lost before the temperature dependence of the susceptibility deviates significantly from mean-field form. This presents a challenge to ``general theories'' of pseudogap\cite{mueller_review_2017}, since these two phenomena would need to occur in the same order with respect to changes of the system parameters (e.g. interaction strength, doping) across the various systems where such a theory might be applied. Finally, we mention that the method we introduced to measure the susceptibility is versatile, and can be feasibly extended to various ultracold atom systems, for example spinor Bose gases, 2D gases, or gases within optical lattices.

\acknowledgements{
We acknowledge support from the Air Force Office of Scientific Research, Young Investigator Program, through grant number FA9550-18-1-0047
}

\bibliographystyle{apsrev4-1}
\bibliography{reference1}

\end{document}